\documentstyle[11pt,paspconf]{article}
\newcommand\simlt{\lower.5ex\hbox{$\; \buildrel < \over \sim \;$}}
\newcommand\simgt{\lower.5ex\hbox{$\; \buildrel > \over \sim \;$}}

\begin{document}

\title{Corona Energy Budget in AGN and GBHCs}

\author{Sergei Nayakshin} 

\affil{Department of Physics, the University of Arizona, Tucson, AZ, 85721}

\begin{abstract}
Recent progress in observations and understanding of spectra of
Seyfert Galaxies shows that X-rays are most likely produced by
magnetic flares on the surface of the disk, similar to Solar X-ray
emission. However, while the model reproduces the shape of the
spectrum well, the question of the overall normalization of the X-ray
component relative to the bolometric luminosity has not been
previously considered. Here we show that, in gas-dominated accretion
disks, the magnetic energy transport can indeed power the corona in a
way consistent with observations, {\it if} magnetic field in the disk
is mostly contained to strong magnetic flux tubes. However, in
radiation dominated disks radiation diffusion makes the field
weak/diffuse, and thus the magnetic energy transport is less
efficient, i.e., in such disks the disk intrinsic emission should be
the dominant component in the overall spectrum. We compare our
findings to observations, and conclude that our theory can account for
the often observed ``steeper when brighter'' behavior in AGN and the
hard-soft spectral transitions in GBHCs.
\end{abstract}

\keywords{Magnetic fields, accretion disks}

\section{Introduction}

The two-phase patchy corona-disk model (PCD model hereafter) was
suggested by Haardt \& Maraschi (1991,1993) \& Haardt et al. (1994) to
explain the almost unique X-ray spectral index of Seyferts. Via
accurate spectral calculations, Stern et al. (1995), Poutanen \&
Svensson (1996) \& Poutanen et al. (1997) showed that the model
naturally explains observed distribution of the X-ray spectral
indexes, strength of the fluorescent iron line, and the presence of
Compton reflection component for Seyfert Galaxies (see also Svensson
1996).

The basic physics of the PCD model is well understood qualitatively
(e.g., Haardt et al. 1994, Galeev et al. 1979, Nayakshin 1998b),
because it is adapted after Solar magnetic flares (e.g., Parker 1979,
Priest 1982, Tsuneta 1996, Tajima \& Shibata 1997). Namely, turbulent
accretion disks are expected to produce magnetic fields very
efficiently, since Keplerian accretion disks are differentially
rotating strongly ionized plasmas. The magnetic fields tend to
concentrate into magnetic flux tubes -- regions of enhanced magnetic
field, with magnetic pressure comparable to the ambient gas pressure
(Parker 1979, Chapter 10). Because the flux tubes immersed in the
fluid must be in pressure equilibrium with that fluid, the gas density
inside the tubes is lower than the ambient gas density. Parker (1955)
has shown that the tubes are therefore buoyant. Thus, these tubes are
expelled out of the disk. Once above the disk, the magnetic fields of
these tubes can reconnect -- transfer their energy to particles
trapped within the tubes, producing hot active regions, or ``patches''
in the patchy corona model.  These active regions then produce X-rays
by inverse Compton upscattering of the disk radiation and by other
emission mechanisms.

The broad-band spectrum of accretion disk in this picture consists of
basically three components: (1) the X-rays produced by magnetic flares
in the disk corona; (2) radiation due to reprocessing of these X-rays
in the disk below the flares (which we think is the origin of the Big
Blue Bump of Seyfert 1s [BBB hereafter; Nayakshin \& Melia 1997,
Nayakshin 1998b]); (3) the internal optically thick disk
emission. Following Svensson \& Zdziarski (1994), let us parameterize
the fraction of the overall disk power released through the flares as
$f< 1$. If the disk bolometric luminosity is $L$, the intrinsic disk
emission then accounts for luminosity $(1-f) L$. Since roughly half of
the X-rays produced by the flares are directed back to the disk (e.g.,
Haardt et al. 1994), the observed X-rays and the BBB should both
produce luminosities $L_x \simeq L_{bbb} \simeq (1/2) f L$.  Here we
attempt to address the magnitude of $f$ in the context of the PCD
model, and compare it to observations of both AGN and GBHCs.

\section{Energy Transfer by Buoyant Magnetic Fields}

Let us estimate the magnetic energy transport due to rising magnetic
flux tubes and compare it with the total disk emission (energy
liberated per unit area of the disk).  The latter can be shown to be
(e.g., Frank et al. 1992, Chapters 4,5) $F_{\rm tot} \simeq \alpha c_s
P_{\rm tot}$, where $\alpha$ is the Shakura-Sunyaev viscosity
prescription, $c_s$ is the sound speed in the mid-plane of the disk,
and $P_{\rm tot}$ is the total (gas plus radiation) pressure in the
disk. The time averaged magnetic energy flux is simply given by
$F_{\rm mag} \simeq v_b \langle{P_{\rm mag}}\rangle$, where $v_b$ is
the average buoyant rise velocity, which cannot be larger than $c_s$,
and $\langle{P_{\rm mag}}\rangle$ is the volume average of the
magnetic field pressure. Thus, the importance of the magnetic energy
transport depends on the ratio $\langle{P_{\rm mag}}\rangle/P_{\rm
tot}$. The viscosity parameter was defined by Shakura \& Sunyaev
(1973) to be
\begin{equation}
\alpha \simeq \nu_t + {\langle{P_{\rm mag}}\rangle\over P_{\rm tot}} ,
\label{ssvis}
\end{equation}
where the first term, $\nu_t$ is the turbulent viscosity. Thus,
$\langle{P_{\rm mag}}\rangle\simlt \alpha P_{\rm tot}$, which yields
\begin{equation}
f\equiv {F_{\rm mag}\over F_{\rm tot}} \simlt {v_b\over c_s}\simlt 1
\label{eq1}
\end{equation}
The buoyant rise velocity depends on properties of the magnetic flux
tubes and the gas around them. For flux tubes of size $a_0\ll H$,
where $H$ is the disk height scale, Vishniac (1995a,b) shows that
$v_b\sim \alpha c_s$, which makes $f$ a small number. Physically, the
problem arises because during its rise to the disk surface, magnetic
field also contributes to the angular momentum transfer in the disk,
which in turn leads to further liberation of energy in the disk. This
latter ``side'' process in fact produces more heat than magnetic
buoyancy itself, and so it would appear that the PCD model should
always produce more (and much more, if $\alpha\ll 1$) thermal disk
emission than the corona can, which is at odds with observations.  The
strongest constraint here comes from GBHCs in their hard state, whose
spectra contain most of the power in the hard power law, so that $f$
is probably $\simeq 0.8$. For Seyferts, the X-ray emission often
accounts for a significant fraction of the overall emission, and $f$
as large as $0.5$ seems to be needed.

A possible solution to this dilemma lies in the fact that estimate
$\alpha\sim\langle{P_{\rm mag}}\rangle/P_{\rm tot}$ is only correct
for a diffuse magnetic field, and a similar argument carefully applied
to a field localized to strong magnetic flux tubes shows that these
fields can contribute considerably less to the disk viscosity and thus
the local heating.  The reduced local heating would then explain how
it is possible for real accretion disks to have $f$ approaching unity.
Our point here is that a diffuse and tangled magnetic field of
sub-equipartition intensity will be simply carried along with the
fluid, i.e., it will take part in the differential rotation of the
fluid. By resisting the stretching through magnetic field tension, the
field will contribute to viscosity in the usual way. At the same time,
a flux tube is an entity of its own, which manifests itself in the
fact that the tube can move with respect to the fluid, e.g., be
buoyant. Accordingly, the flux tube may avoid the stretching by simply
not following these motions of the fluid that try to deform the
tube. Thus, the flux tube may contribute to viscosity at a smaller
rate than a diffuse field would do.

Suppose for simplicity the shape of the tube is that of a torus with
the larger radius $a_0\simlt H$ and the smaller radius $a\leq 2
a_0$. There is a viscous drag force $D$ on the flux tube in this case,
caused by the friction as the fluid flows by the tube:
\begin{equation}
D \simeq C_d \, \rho\, v_d^2\, a \, a_0/2
\label{drforce}
\end{equation}
(e.g., Parker 1979, \S 8.7, and references there), where $C_d$ is the
dimensionless drag coefficient, $\rho$ is the gas density and $v_d$ is
the differential flow velocity, which is $v_d\sim c_s a_0/H$ for a
Keplerian accretion disk. For the flux tube not to be deformed by the
drag force, the tube magnetic tension $T$ should exceed this force:
\begin{equation}
T \simeq P_{\rm mag} 2 \pi a^2 > D
\label{tens}
\end{equation}
The ratio of these two forces is
\begin{equation}
{T\over D} \sim {4\pi a\over a_0} C_d^{-1} {P_{\rm mag}\over P_{\rm
tot}} \left (H\over a_0\right )^2
\label{tod}
\end{equation}
where we used $P_{\rm tot}= \rho c_s^2$. The value appropriate for the
drag coefficient in accretion disks is $C_d\sim 1/4$ (following
Vishniac 1995, Stella \& Rosner 1984, Sakimoto \& Coroniti 1989,
Parker 1979). Thus, equation(\ref{tod}) asserts that for flux tubes
with magnetic field pressure comparable to the equipartition value,
and the size $a_0$ smaller than the disk scale hight $H$, $T > D$, and
thus the tubes cannot be deformed by the flow in this case, and
instead are dragged around almost as a solid body. The contribution of
the flux tube to the angular momentum transfer is reduced by the same
factor $\sim D/T$, since the flux tube is being stretched at a rate
slower than implied by the differential flow of the ambient gas around
the tube. If all the magnetic field is in the form of strong flux
tubes for which the magnetic tension exceeds the drag force, then the
limits on the magnetic field volume average become
\begin{equation}
\langle {D\over T} P_{\rm mag}\rangle \simlt \alpha P_{\rm tot}
\label{fod}
\end{equation}
We can now estimate the ratio of the magnetic energy
flux $F_{\rm m}$ to the radiation energy flux as
\begin{equation}
{f\over 1-f}\, \simeq {v_b\over c_s}\, {1 + T\over D}\,{\rm ,}
\label{magflux}
\end{equation}
which is much easier to reconcile with the magnetic energy flux
requested by the two-phase corona-accretion disk model, since now the
buoyant rise velocity can be comfortably below its absolute maximum
value, i.e., the sound speed $c_s$ and yet provide magnetic energy
flux exceeding the radiation flux.

A simple physical analogy here is that of a sail on a ship. When the
sail is ``on'', the force (due to wind) acting on the sail is many
times larger than it is in the case of the sail that is folded in. The
amount of this wind-sail interaction clearly depends not on the
overall mass of the sail, but on the state of the sail -- whether it
is open and positioned properly with respect to wind or whether it is
rolled in a tube. Similarly, with same volume average magnetic field
one gets less or more interaction between differential flow and the
field depending on whether the field is uniform in space, or is in
strong flux tubes, such that most of the flow simply miss the tubes to
interact with them.

\section{Radiation Pressure and Properties of a Single Flux Tube}
\label{sect:radp}

Considering magnetic fields in the previous section, we did not
explicitly separate the total pressure $P_{\rm tot}$ into the
radiation pressure $P_{\rm rad}$ and the gas pressure $P_{\rm
gas}$. Our initial neglect of the radiation pressure dynamical effects
is equivalent to the assumption that radiation and particles move
together, as one fluid.  Such approach is valid as long as scales of
interest are much larger than the photon mean free path, since in this
case the radiation is essentially ``glued'' to particles due to the
large opacity. In the opposite limit of small opacity, the gas and
radiation will behave very differently with respect to magnetic
fields. Whereas radiation does not directly interact with magnetic
field (i.e., these two components are unaware of each other's
presence), particles in ideal MHD cannot cross magnetic field lines
and their motion is constrained to the direction parallel to the
field. Let us then compare the time scale for the radiation diffusion
into the flux tube with a time scale important for generation and
maintenance of strong magnetic flux tubes. The radiation diffusion
time scale $t_{\rm d}$ can be estimated as $t_{\rm d}\sim (a/c) n'_e
\sigma_T a$, where $n'_e$ is the particle density inside the flux
tube, which we can assume to be of the order of the disk particle
density $n_e$, and $\sigma_T$ is the Thomson cross section.

Turbulent motions of the fluid are believed to be the mechanism for
the magnetic field amplification (e.g., Vishniac 1995a,b). Let $u_{\rm
t}$ be the typical turbulent velocity, and $\lambda_{\rm t}$ be the
turbulent length scale (corresponding to the largest eddy length
scale).  The gas executes turbulent motions on the eddy turn over time
scale $t_{\rm t}\equiv \lambda_{\rm t}/u_{\rm t}$. This yields
\begin{equation}
{t_{\rm d}\over t_{\rm t}} \sim {a^2\over H \lambda_{\rm t}}\,
{u_{\rm t}\over c_s} {\tau_d c_s\over c}
\label{dift}
\end{equation}
where $\tau_d$ is the disk Thomson optical depth, and $a\simlt
\lambda_{\rm t}$ (Vishniac 1995a). Further, in the standard
Shakura-Sunyaev viscosity prescription, the turbulent velocity and
spatial scale are parameterized by $u_{\rm t} \lambda_{\rm t} = \alpha
c_s H$. Finally, in the radiation pressure dominated region of the
disk, the standard disk equations lead to $\tau_d c_s/c\simeq
\alpha^{-1}$, for arbitrary radii and accretion rate. Therefore, one
can see from Equation (\ref{dift}) that the ratio of the diffusion
time scale to the turbulent time scale is of the order
unity. Moreover, we compared $t_{\rm d}$ with {\it one} eddy turn over
time scale, whereas generation of the field comparable with the
equipartition value is likely to take much longer time, simply because
one turbulent eddy does not carry enough energy (we assume that
turbulence is sub-sonic). Due to this the diffusion of radiation into
the flux tubes is much faster than the field generation process.

As a result of the efficient radiation diffusion, the radiation
pressure inside a flux tube should be equal to the ambient radiation
pressure, which means that the magnetic field pressure of the tube can
only be as large as the gas pressure $P_{\rm gas}$. A simple analogy
here is a car tire that is pumped with a gas that easily diffuses in
or out.  Obviously, difference in the gas pressure inside and outside
will be small which will render such a tire useless. In the case of
the flux tubes, $P_{\rm mag}\simlt P_{\rm gas} \ll P_{\rm tot}$ for
the radiation-dominated disks means that the tubes are now weak and
easily stretched by the differential motions of the disk, and thus
$\langle{P_{\rm mag}}\rangle\sim \alpha P_{\rm tot}$. The tension
force is now small compared with the drag force due to differential
Keplerian flow, and, according to equation (\ref{magflux}), $f/(1-f)
\simeq v_b/c_s < 1$. In physical terms, the X-ray luminosity should
always be smaller than the optical- to soft X-ray luminosity in the
radiation dominated accretion disks.

\section{Spectral States of AGN and GBHCs}\label{sect:states}

In order for magnetic flares to be in the parameter space of the PCD
model, the magnetic fields in the active regions should be sufficiently
strong to make the compactness parameter $l\gg 0.01$, and the disk
intrinsic flux should be much smaller than the X-ray flux. Since the
field is limited to the equipartition value in the disk, one finds
$l\propto \dot{m} \alpha^{-1} (P_{\rm mag}/P_{\rm tot})$ (e.g.,
Nayakshin 1998b). Numerically, it seems unlikely that accretion disks
dimmer than $\dot{m}\simlt 10^{-4}$ or so will be able to produce
flares of the right compactness. Thus, we believe that to produce the
typical hard X-ray spectrum, AGN must accrete above some minimum
accretion rate $\dot{m}_d \sim 10^{-4}$.

The transition from the gas to radiation -dominated disks happens at
$\dot{m} = \dot{m}_r$:
\begin{equation}
\dot{m}_r = 2.2 \times 10^{-3} \left( \alpha M_8\right)^{-1/8}\,
(1-f)^{-9/8}
\label{mr}
\end{equation}
For accretion rates $\dot{m}_d < \dot{m} < \dot{m}_r$, the magnetic
flares satisfy PCD model constraints, and thus their X-ray spectra
should be hard, (i.e., typical hard Seyfert spectrum for AGN and
typical hard state spectrum for GBHCs). In addition, since magnetic
buoyancy can transfer more energy out of the disk than the usual
radiation flux, most of the power is contained in the X-ray power law
(i.e., $f\simeq 1$) {\it and} the emission due to reflection and
reprocessing of this radiation in the disk.

As the accretion rate increases above $\dot{m}_r$, the importance of
X-ray production by magnetic flares decreases, i.e., the fraction $f$
is decreasing as $\dot{m}$ increases. This means that intrinsic disk
emission becomes the dominant feature in the spectrum of an AGN or a
GBHC. Note that the expression for $\dot{m}_r$ given by equation
(\ref{mr}) depends on $f$ itself, but since radiation-dominated disks
are not effective in transporting energy into the corona, factor $1-f$
should not be too small for such disks. In fact, if we use
observational constraints on $f$ from hard/soft spectral transitions
in GBHCs, $1-f\simeq 1/2$ at $\dot{m}=\dot{m}_r$ (see \S 5).

Based on theoretical arguments alone, we cannot be certain about what
happens to the {\it shape} (as opposed to already discussed {\it
normalization}) of the X-ray spectrum from flares when $P_{\rm rad}\gg
P_{\rm gas}$. The uncertainty is present due to our ignorance of the
numerical value of $\alpha$ when radiation pressure exceeds the gas
pressure.  The standard accretion disk theory in this case is unstable
to viscous and thermal perturbations (e.g., Frank et al. 1992), and so
the form of viscosity law in radiation dominated disks remains a
highly controversial issue. However, if we assume that the viscosity
is proportional to the gas pressure only (for motivation see Lightman
\& Eardley 1974, Stella \& Rosner 1984, Sakimoto \& Coroniti 1989),
the accretion disk is stable, and it turns out that the decrease in
the effective value of $\alpha$ may compensate for the fact that
magnetic fields are limited to the gas pressure, so that the flares
may still be in the PCD model parameter space. Namely, the effective
$\alpha$ scales as $\alpha_g P_{\rm gas}/P_{\rm tot}$ in this case,
where $\alpha_g\leq 1$ is a constant. Thus, $l\propto (\dot{m}/\alpha)
\,(P_{\rm gas}/P_{\rm tot}) = (\dot{m}/\alpha_g)$, that is, the
compactness parameter does not have to decrease as the disk goes from
the gas- to the radiation-dominated regime.

Depending on behavior of $\alpha$ in the radiation-dominated disks,
the X-ray spectrum from magnetic flares will either be unchanged from
that of the hard state, or will become steeper because of the extra
cooling of the active regions caused by the stronger intrinsic disk
emission. The X-ray spectrum will steepen when the disk intrinsic flux
will approach the X-ray flux from a flare. This happens for accretion
rates above $\dot{m}_{\rm soft}$, where $\dot{m}_{\rm soft}$ is given
by
\begin{equation}
\dot{m}_{\rm soft}\, = \, 0.06 \,(l/0.1)^{1/2} \left(1-f_s\right)^{-1}
\label{dms}
\end{equation}
As in equation (\ref{mr}), $f_s$ is the fraction of power reprocessed
via magnetic flares and depends on the accretion rate itself, but we
again expect that $1-f_s\simeq 1$. Further, we scaled the poorly known
compactness parameter on $0.1$, since X-ray reflection calculations
(Nayakshin 1998a,b) point to $l$ of this order in both AGN and GBHCs.

Thus, the theory predicts that the X-ray spectrum should become
steeper when $\dot{m}$ increases above $\dot{m}_{\rm soft}\sim
0.06$. For even higher accretion rates ($\dot{m} \simgt 0.2$ or so),
the disk becomes geometrically thick, and advection of energy (in the
sense of Abramowicz et al. 1988) in the black hole will change the
properties of the disk further, so that we do not attempt to describe
those disks due to theoretical uncertainties. We can then complete
classifying accretion disks by naming the radiation-dominated state
with $\dot{m}_r\simlt \dot{m}\simlt \dot{m}_{\rm soft}$ the
``intermediate'' one, the state with $\dot{m}_{\rm
soft}\simlt\dot{m}\simlt 0.2$ ``soft'', and, finally, the one with
$\dot{m}\simgt 0.2$ ``very high'' in analogy with the very high state of
GBHCs.

\section{Comparison With Observations}\label{sect:obs}

\subsubsection{Typical Hard Seyferts.}  As discussed in \S 4, our 
model predicts that Seyferts with the typical hard X-ray spectra
should accrete at accretion rates below $\dot{m}_{\rm soft}$. In order
to estimate $\dot{m}\equiv L/L_{\rm Edd}$ based on observations, we
need to know AGN masses. Although these are unknown, variability
studies may provide some help. For example, the global compactness
parameter $l_g$ has been estimated for a sample of Seyfert Galaxies by
Done \& Fabian (1989). In their estimate, they assumed that the
typical size of the emitting region is given by the distance traveled
by light during the shortest doubling time scale $\Delta T$ observed
for a given source ($l_g= \sigma_T L_x/(m_e c^4 \Delta T)$) . For an
accretion disk this typical size should be of order $\sim 10
R_g$. This yields $\dot{m} \simeq (10 m_e/ 2\pi m_p) \, l_g/2 \,
\simeq l_g/2000$. (see also Fabian 1994). Now, in Table 1 of Done \&
Fabian (1989), the maximum compactness is about 200, thus maximum
$\dot{m}\sim 0.1$. Moreover, $80$\% of the sample have $\dot{m} <
0.02$, with smallest values of the order of $10^{-4}$.  These
estimates do not include the BBB, which could make a significant
contribution to the bolometric luminosity of Seyfert
Galaxies. However, using $1375$ Angstrom fluxes reported by Walter \&
Fink (1993) for the sources with the highest values for the
compactness, we estimated $L_{\rm uv} \sim L_{\rm x}$, so that
inclusion of the emission at lower wavelengths did not affect our
conclusions significantly.

Sun \& Malkan (1989) fitted multi-wavelength continua of quasars and
AGNs with improved versions of standard accretion disk models. They
found that low-redshift Seyfert Galaxies radiate at only few percent
of their Eddington luminosities. Rush et al. (1996) studied soft X-ray
(0.1-2.4 keV) properties of Seyfert Galaxies. Their results indicate
that $\sim 90$\% of sources in their sample have soft X-ray luminosity
below $10^{44}$ erg/s (with the mean value of order $\sim 10^{43}$
erg/s). If we assume the typical Seyfert 1 spectrum above 2.4 keV,
i.e. a power law with intrinsic photon index $\simeq 2$ and the cutoff
at several hundred keV (e.g., Zdziarski et al. 1996), then total
X-ray/gamma-ray luminosity of these objects can be a factor of 2-3
higher than the soft X-ray luminosity.  Nevertheless, if a typical
Seyfert Galaxy has the black hole mass of $\sim 10^8$, then the average
bolometric luminosity of the Rush et al. (1996) sample is at or below
$\sim 1$\% of the Eddington luminosity. For NGC5548, using results of
Kuraszkiewicz, Loska \& Czerny (1997), we obtained $\dot{m}\sim (4-16)
\times 10^{-3}$. Summarizing, there is some evidence that X-ray hard
Seyfert 1 Galaxies accrete at a relatively low accretion rate, i.e.,
from probably just below $\dot{m} = 0.1$ to very low accretion rates
of $\sim 10^{-4}$.

\subsubsection{Steep X-ray spectrum AGN, Narrow Line Seyfert 1 (NLS1) 
Galaxies.} Relatively recently, it was found that a subset of Seyfert
Galaxies have unusually steep soft X-ray spectra (for a review, see
Pounds \& Brandt 1996, PB96 hereafter, and Brandt \& Boller 1998).
Common properties of the group include steep spectra, rapid
variability, strong Fe II emission and identification with NLS1. PB96
speculated that the most likely explanation for the steep X-ray
spectrum is an unusually high accretion rate. PB96 also showed that
soft X-ray (i.e., 0.1-2 keV) spectral index is strongly correlated
with the width of the H$\beta$ line for a sample of Seyfert Galaxies.
Wandel \& Boller (1998) and Wandel 1998 suggested an explanation of
the correlation based on the simple idea that steeper X-ray spectrum
implies a larger ionizing UV luminosity, which translates into a
larger broad line region size, and thus a smaller velocity dispersion
(since Keplerian velocity is $\propto 1/R^{1/2}$). They found that
masses of the narrow-line Seyfert Galaxies tend to be lower that those
of typical broad H$\beta$ line Seyferts, and thus have larger
$\dot{m}$. Brandt, Mathur \& Elvis (1997) found that the higher energy
{\it ASCA} slopes (2-10 keV) correlate with the H$\beta$ line as
well. Thus, NLS1 galaxies often have intrinsic X-ray slope that is
steeper than that of normal Seyferts.

Laor et al. (1997) found the same correlation for a sample of quasars,
and suggested that NLS1 galaxies accrete at a higher fraction of the
Eddington accretion rate than normal Seyferts do. They assumed that
the bulk motion of the broad line region is virialized, and that the
scaling of the BLR with luminosity is that found from reverberation
line mapping of AGN (e.g., Peterson 1993).  In this case larger
luminosities $L$ correspond to larger BLR size, and thus smaller
H$\beta$ FWHM. Now, if $\Gamma$ is larger for higher $\dot{m}$, then
the observed relation (smaller H$\beta$ FWHM -- larger $\Gamma$)
ensues.  However, no reason for $\Gamma$ to become larger with
increasing $\dot{m}$ was given, except for a heuristic suggestion of
Pounds et al. (1995) that if the power released in the corona remained
constant, then the X-ray index would become steeper with increasing
bolometric luminosity (which is equivalent to assuming $f/(1-f)
\propto (\dot{m})^{-1}$).  Our theory of accretion disk states may
provide a natural explanation for this spectral steepening (see \S 4).

\subsubsection{Observational Constraints on Corona/Disk Energy 
Partitioning.} So far in \S 5, we discussed how the shape of the X-ray
spectrum depends on $\dot{m}$.  Since we believe that the BBB is
produced by reflection of X-rays produced by the flares off the
surface of the disk, we should also see if we can test our theory
based on observations of BBB. Here we will only concentrate on the
overall normalization of the bump, taking up the question of its
spectrum in the other contributed paper in these proceedings
(Nayakshin 1998a).

Walter \& Fink 1993, Walter et al. 1994 and Zhou et al. 1997 studied
the BBB of Seyfert 1's and found that this spectral feature is
ubiquitous in these sources. The observed spectral shape of the bump
component in Seyfert 1's hardly varies, even though its luminosity
$L_{\rm bbb}$ ranges over 6 orders of magnitude from source to
source. However, recent work of Zheng et al. (1997) and Laor et
al. (1997) showed that quasars in their (different) samples do not
show the BBB component.

There are two reasons why we think that there is no inconsistency
here.  As discussed above, typical Seyfert Galaxies should accrete at
a relatively small accretion rates, i.e., $\dot{m}\simlt \dot{m}_{\rm
soft}$, which corresponds to luminosity $L \simeq 5 \times 10^{44}
M_8$ ergs/sec. In the sample of Walter \& Fink (1993), very few
objects have UV luminosity above few $\times 10^{44}$ ergs/sec,
whereas Zheng et al. (1997) fit to the mean spectrum in their sample
gives $L\simeq 8.5 \times 10^{45}$ ergs/sec. Similarly, almost all AGN
in Laor et al. (1997) sample have $L_{3000}> 10^{45}$
ergs/sec. Therefore, Walter \& Fink (1993) sample contains Seyferts
that are dimmer than sources in the other two samples by factors 10 to
100. Accordingly, sources in Walter \& Fink (1993) sample may accrete
at hard or intermediate state $\dot{m} \simlt \dot{m}_{\rm soft}$,
whereas AGN of the two other samples could accrete at the soft and/or
very high state regime. Our first argument is that according to
discussion in \S 4, the more luminous sources must be X-ray weak,
i.e., $f\ll 1$ and thus the normalization of the BBB is $\sim f/2\ll
1$, so that it disappears on the background of the disk thermal
emission ($1-f\simeq 1$) in samples of Zheng et al. (1997) and Laor et
al. (1997). Our second argument has to do with the shape of the
reprocessed spectrum. Namely, we find (Nayakshin 1998a,b) that an
ionization instability may lead to the reprocessed spectrum being a
power law with basically same index as that of the incident X-rays, up
to $\sim 30 keV$, which will blend to undetectability with the X-ray
spectrum from the flares (see also the limit of the ``hot medium'' in
Zycki et al. 1994).

Another well known observational fact for quasars is the correlation
between the optical to X-ray spectral slope $\alpha_{ox}$ and the
optical luminosity (Green et al. 1995). The optical to X-ray index
$\alpha_{ox}$ is not a real spectral index in this energy range, but
is defined as the index of an imaginary power law connecting observed
optical and X-ray emission.  Wilkes et al. (1994), Green et al. (1995)
show that more luminous sources have larger $\alpha_{ox}$, i.e., more
luminous objects have comparatively less X-ray emission. This is again
in a qualitative agreement with our theory.

\subsubsection{GBHCs state transitions} As Nayakshin (1998a) shows, 
PCD model with magnetic flare parameters (most notably $l$) same as
for AGN case, when re-scaled for the case of GBHCs gives harder X-ray
spectra and weaker reprocessing features (as compared to AGN),
explaining peculiarities of hard state spectra of GBHCs such as
Cyg~X-1. We thus will try to use the same logic as for AGN, only
changing $M$ from $\sim 10^8$ to $\sim 10$, to understand GBHCs
spectral states, that are often referred to (in order of increasing
$\dot{m}$) as hard, intermediate, soft and ultra-soft state (e.g.,
Grove, Kroeger \& Strickman 1997 and Grove et. al. 1998). Note that
for $M\sim 10 \,M_{\odot}$, the transition from the gas- to
radiation-dominated regimes happens at a considerably higher
$\dot{m}$, namely, $\dot{m}_r\simeq 1.7\times
10^{-2}\,(1-f)^{-9/8}$. Further, a transition is in fact defined by
fraction $f$ changing from large to small numbers, so we can assume
$f\simeq 1/2$, such that $\dot{m}_r\simeq 0.037$. This number is
remarkably close to where the state transition are observed to occur
for GBHCs.

\begin{figure*}[t]
\plotfiddle{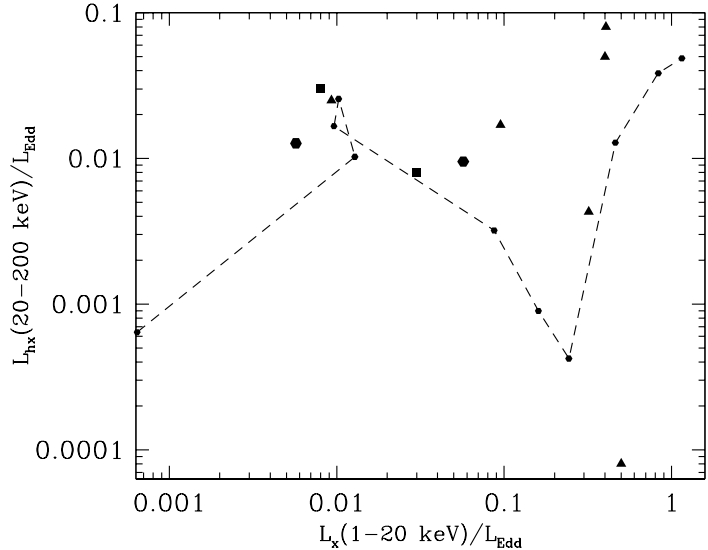}{150pt}{0}{50}{50}{-160}{-160}
\caption{Division of power between hard X-ray luminosity $L_{\rm hx}$
(20 -- 200 keV) and soft X-ray luminosity $L_{\rm x}$ (1 -- 20 keV)
for GBHCs. Note that most of the luminosity is in the hard component
for $\dot{m}\simlt 0.04$. See text and Nayakshin (1998b) for details.}
\label{fig:gbhc}
\end{figure*}

Barret, McClintock \& Grindlay (1996) assembled a sample of GBHCs in a
$L_{\rm hx}-L_{\rm x}$ phase space, where $L_{\rm hx}$ is the hard
X-ray luminosity in the range $20-200$ keV, and $L_{\rm x}$ is the
X-ray luminosity in the range $1-20$ keV. One of the striking results
of this exercise is that GBHCs always have most of their power in the
soft X-ray component (presumably the optically thick disk emission)
when they radiate at a high fraction of their Eddington luminosity.
Barret et al. (1996) also plotted the evolutionary track of the GBHC
transient source GRS 1124-68 on the same $L_{\rm hx}-L_{\rm x}$ phase
diagram. Since we are interested in the dimensionless accretion rate,
we reproduce data of Barret et al. in terms of $L_{\rm x}/L_{\rm Edd}$
and $L_{\rm hx}/L_{\rm Edd}$ in Figure (1), adding some data for
Cyg~X-1.  Figure (1) shows that the spectra of GBHCs indeed have most
of their power in the hard component up to $\dot{m} \sim 0.04$, and
then there is a rather strong spectral transition, which confirms our
finding that $f$ decreases with increasing $\dot{m}$ for $\dot{m} \geq
\dot{m}_r$. This behavior unites AGN and GBHCs.

There are differences, too. Just as in AGN case, above the gas- to
radiation transition, our theory predicts existence of the
intermediate state ($\dot{m}_r\leq \dot{m}\leq\dot{m}_{\rm
soft}$). However, note that for GBHCs the intermediate state is
squeezed in in the narrow interval between the hard and the soft
states, whereas for AGN it is not the case because $\dot{m}_r$ is much
lower than $\dot{m}_{\rm soft}$ (if compactness parameter $l$ is
indeed of order $\sim 0.1$ in both AGN and GBHCs, as suggested by
spectral modelling). The difference is then such that the intermediate
state essentially disappears in GBHCs, so that as $f$ decreases at the
region $\dot{m} \sim \dot{m}_r$, the X-ray spectral index steepens,
while for AGN case it still may stay hard until $\dot{m}$ reaches
$\sim \dot{m}_{\rm soft}$.

\section{Discussion}

We have shown that one cannot power coronae of accretion disks with
diffuse magnetic fields, at least not in a way consistent with
observations of hard states in Seyferts and GBHCs. In short, the
problem is that these fields are transported into the corona ``too
slowly'' ($v_b$ is realistically just a fraction of sound speed), so
that during their rive, the fields contribute ``too much'' to the
angular momentum transport in the disk and to the local disk heating,
which in turn produces disk thermal radiation. It is then not possible
to ever produce more power in X-rays than in the disk thermal emission
(i.e., $f\ll 1$), which is inconsistent with observations.

However, if one assumes that most of the disk magnetic field is
localized in the form of strong magnetic flux tubes, the field
contribution to viscosity goes down substantially. This then leads to
a decrease of the local disk heating, so that it becomes possible for
accretion disks to release most of their energy in coronae rather than
through the common thermal radiation diffusion. 

We further have shown that in radiation-dominated accretion disks,
diffusion of radiation into flux tubes does not allow the tube
magnetic fields to reach equipartition values, and that instead the
field pressure is limited to the gas pressure. Thus, we concluded that
magnetic fields in the radiation-dominated disks are in the diffuse
form, and thus such disks cannot produce X-rays as efficiently as the
gas-dominated disks can. Using this idea, and some spectral
constraints from PCD model, we also showed that the observed spectral
transitions and states of GBHCs and the mounting evidence for presence
of similar states in AGN accretion disks support the theory of
accretion disks with magnetic flares. We believe that this, together
with other successes of the theory, provides one with optimism that
magnetic flares is the physics that was missing in the standard
accretion disk theory for decades.

\acknowledgments The author is very thankful to the workshop
organizers for the travel support, and to F. Melia for support and
useful discussions in an early stage of this work.

\end{document}